\documentclass[twocolumn]{jpsj3ff}
\usepackage{txfonts}

\title{
Fractional skyrmion and absence of low-lying Andreev bound states
in a micro fractional-flux quantum vortex
}

\author{Takashi Yanagisawa, Yoichi Higashi, and Izumi Hase}

\inst{
Electronics and Photonics Research Institute,
National Institute of Advanced Industrial Science and Technology,
1-1-1 Umezono, Tsukuba, Ibaraki 305-8568 Japan
}

\abst{
We investigate quasi-particle excitation modes and the topological number 
of a fractional-flux
quantum vortex in a layered (multi-component) superconductor.
The Bogoliubov equation for a half-flux quantum vortex is solved to show that 
there is no low-lying
Andreev bound state near zero energy in the core of a quantum vortex, 
which is surprisingly in
contrast to the result for an inter-flux vortex.  Related to this result,
there are singular excitation modes that have opposite angular momenta,
moving in the opposite direction around the core of the vortex.
The topological index (skyrmion number) for a fractional-flux quantum
vortex becomes fractional
since the topological index is divided into two parts where one from the
vortex (bulk) and the other from the kink (domain wall, boundary).
The topological numbers for both the vortex and the kink (domain wall) are fractional,
and their sum becomes an integer.
This shows an interesting analogy between this result and the index theorem for 
manifolds with boundary.
We argue that fractional-flux quantum vortices are not
commutative each other and follow non-abelian statistics.
This non-abelian statistics of vortices is different from that in
p-wave superconductors.
}

\begin{document}
\maketitle

\section{Introduction}

Multi-component superconductivity has been studied 
intensively\cite{mos59,suh59,per62,kon63,sta10,tan10a,tan10b,dia11,yan12,hu12,
sta12,pla12,mai13,wil13}.
There appear many significant phenomena in multi-component superconductors
such as time-reversal symmetry breaking\cite{sta10,tan10a,tan10b,dia11,yan12,
hu12,sta12,pla12,mai13,wil13,gan14,yer15}, the appearance of massless 
modes\cite{yan13,lin12,kob13,koy14,yan14,tan15,sha02,yan17c}, the existence of
fractional-flux quantum vortices\cite{yan12,izy90,vol09,tan02,bab02,col05,blu06,gor07,
chi10,kup11,gar11,pin12,gar13,smi05,tan18,hay19,nag18},
and unconventional isotope effect\cite{cho09,shi09,yan09}.
A fractional-flux quantum vortex (FFQV) has been
examined in the study of multi-component superconductors.
The existence of an FFQV has been proposed theoretically in multi-component
superconductors and there have been
several experimental attempts to observe FFQVs.
Recently, the observation of an FFQV in an Nb thin film 
superconducting (SC) bilayer has been
reported as a magnetic flux distribution image by using a scanning
superconducting quantum interference device (SQUID) microscope\cite{tan18}.
It is then important to clarify quasi-particle excitation modes in an FFQV and
a half-flux quantum vortex (HFQV) in multi-component or layered superconductors.

An FFQV has the fractional vorticity $Q$ where the magnetic flux in an FFQV
is $Q\phi_0$ for the unit quantum flux $\phi_0=h/2|e|$ (in MKS units).
An FFQV can be regarded as a micro vortex ($Q<1$) which is defined against the
giant vortex\cite{vir99}.
We have $Q=1/2$ for an HFQV.
The existence of a fractional-flux quantum vortex is strongly related to
the existence of a kink solution in the phase space of SC
gaps.  In a two-gap superconductor, we have two global phases $\phi_1$ and $\phi_2$
associated with two SC gaps.  The phase difference, defined by
$\varphi= \phi_1-\phi_2$, satisfies the sine-Gordon equation,
$\partial^2 \varphi/\partial y^2-\alpha\sin\varphi =0$,
where $\alpha$ is a constant and we assume that $\varphi$ has spatial dependence
in the direction $y$.
Since the cosine potential is a function of period $2\pi$, there is
a kink solution satisfying the boundary condition that $\varphi\rightarrow 0$
as $y\rightarrow -\infty$ and $\varphi\rightarrow 2\pi$ as
$y\rightarrow \infty$.
An HFQV exists at the end of the kink\cite{yan12,yan18}.  This is because a net-change
of $\phi_1$ is $2\pi$ by a counterclockwise encirclement of the vortex
and that of $\phi_2$ vanishes
by including contributions from the kink.  We can define the total 
vorticity $Q_{T}$ as the net-change of phase $\phi_i$ divided by $2\pi$. 
We have $Q_T=1$ for one band and $Q_T=0$ for the other band.

In this paper, we investigate the quasi-particle excitation modes in an FFQV.
The Bogoliubov equation for an HFQV is solved numerically to obtain the
energy spectra of excitation modes.  We show that there are no quantized excitation 
modes near zero energy
in the core of an HFQV, indicating the absence of low-lying Andreev bound states.
There are quantized excitation modes in the kink as well, which are
Andreev bound states since the gap function changes
its sign when going across the kink.
We also show that the topological number $N_{vortex}$, which is just regarded as the skyrmion
number, becomes fractional for an FFQV: $N_{vortex}=Q$.
The kink also has the fractional topological number.
This is a phenomenon that the part of the index number is absorbed into
the boundary (kink).
Lastly, we argue that two half-flux quantum vortices are not commutative, namely,
they follow non-abelian statistics.

\section{Bogoliubov Equation for an FFQV}

The SC gap function is written as
\begin{equation}
\Delta({\bf r})= \Delta_{0}(r)\exp(-iQ\phi),
\end{equation}
where $\Delta_0(r)= \Delta_{\infty}\tanh(r/\xi_0)$ for $r=|{\bf r}|$
and $Q$ is the vorticity. 
$\xi_0$ is the coherence length defined by $\xi_0=\hbar v_F/\Delta_{\infty}$
with the Fermi velocity $v_F$.
$\phi=\phi(\theta)$ is a function of the
angle variable $\theta$.  In the absence of a kink, we have $\phi(\theta)=\theta$
for all $\theta$.
For an HFQV ($Q=1/2$) in a two-component (bilayer) superconductor, there is the kink 
where the phase $\phi$ changes abruptly.  
The behaviors of $\phi_i(\theta)$ ($i=1,2$) for two gaps as a function of 
$\theta$ are shown in
Fig. 1 and Fig. 2 for the first band and second band, respectively.
In these figures, $\phi_1$ and $\phi_2$ are  shown as a step function for simplicity,
which we call the step function approximation.
We show the phase $\phi_2$ as a function of two-dimensional coordinates
$x$ and $y$ in Fig. 3.
In this paper we consider only one band (one layer) by assuming that the Josephson
coupling is small.
The Bogoliubov equation for an FFQV reads
\begin{eqnarray}
\left(
\begin{array}{cc}
-\frac{\hbar^2}{2m}{\bf \nabla}^2-\epsilon_F  & \Delta({\bf r}) \\
\Delta^*({\bf r})  & \frac{\hbar^2}{2m}{\bf \nabla}^2+\epsilon_F \\
\end{array}
\right) \left(
\begin{array}{c}
u_j({\bf r}) \\
v_j({\bf r}) \\
\end{array}
\right)= \varepsilon_j\left(
\begin{array}{c}
u_j({\bf r}) \\
v_j({\bf r}) \\
\end{array}
\right), \nonumber\\
\end{eqnarray}
where $j$ refers to a quantum number representing a quantum level and  
$\epsilon_F$ indicates the Fermi energy.
We have neglected the vector potential by assuming the high
Ginzburg-Landau parameter.
We write the wave function in the form,
\begin{eqnarray}
\Psi_j({\bf r})&\equiv & \left(
\begin{array}{c}
u_j({\bf r}) \\
v_j({\bf r}) \\
\end{array}
\right)= e^{-iQ\sigma_3\phi/2}e^{i\mu\phi}\left(
\begin{array}{c}
\tilde{u}_{\mu n}({\bf r}) \\
\tilde{v}_{\mu n}({\bf r}) \\
\end{array}
\right) \nonumber\\
&=& \left(
\begin{array}{c}
e^{i(\mu-Q/2)\phi}\tilde{u}_{\mu n}({\bf r}) \\
e^{i(\mu+Q/2)\phi}\tilde{v}_{\mu n}({\bf r}) \\
\end{array}
\right),
\end{eqnarray}
where $\sigma_3$ is the Pauli matrix.
$n$ ia a radial quantum number and $\mu$ denotes the angular momentum 
written as $L_z=\mu\hbar$.
We use the same parametrization as for a giant vortex\cite{vir99}.
In the absence of a kink, $\tilde{u}_{\mu n}$ and $\tilde{v}_{\mu n}$ 
depend only the
radial variable $r$. 

The Bogoliubov equation for particle and hole like excitations is
\begin{eqnarray}
&&\sigma_3\frac{\hbar^2}{2m}\Bigg[ -\frac{\partial^2 f}{\partial r^2}
-\frac{1}{r}\frac{\partial f}{\partial r}
-\frac{1}{r^2}\frac{\partial^2 f}{\partial\theta^2}-k_F^2f \nonumber\\
&& +\left(\mu-\frac{Q}{2}\sigma_3\right)^2\frac{1}{r^2}
\left(\frac{\partial\phi}{\partial\theta}\right)^2f 
-i\left(\mu-\frac{Q}{2}\sigma_3\right)\frac{1}{r^2}
\frac{\partial^2\phi}{\partial\theta^2}f \nonumber\\
&& -2i\left(\mu-\frac{Q}{2}\sigma_3\right)\frac{1}{r^2}\frac{\partial\phi}{\partial\theta}
\frac{\partial f}{\partial\theta} \Bigg]
 +\sigma_1\Delta_0(r)f = \epsilon_{\mu n} f, 
\end{eqnarray}
where we put
\begin{eqnarray}
f=f({\bf r})=f(r,\theta) = \left(
\begin{array}{c}
\tilde{u}_{\mu n}({\bf r}) \\
\tilde{v}_{\mu n}({\bf r}) \\
\end{array}
\right).
\end{eqnarray}

\begin{figure}[htbp]
\begin{center}
\includegraphics[height=6.0cm]{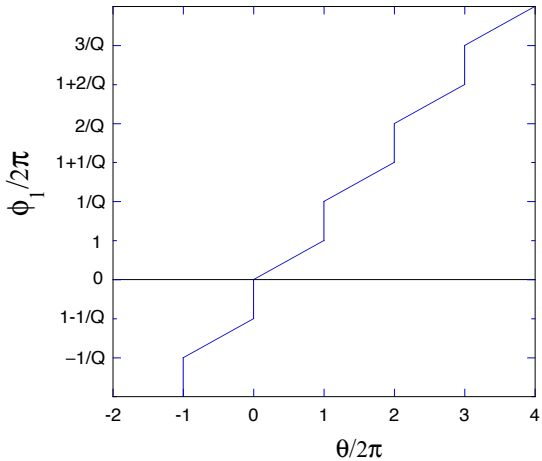}
\caption{ (Color online)
Phase $\phi_1$ of the first band as a function of the angle variable $\theta$
within the step-function approximation.
$\phi_1$ changes at $\theta=2\pi n$ for integer $n$.
$Q$ is the vorticity of a vortex and the phase of the first gap is given by
$Q\phi_1$.  The scale of vertical axis corresponds
to $Q=1/2$.
}
\label{phase1}
\end{center}
\end{figure}

\begin{figure}[htbp]
\begin{center}
\includegraphics[height=6.0cm]{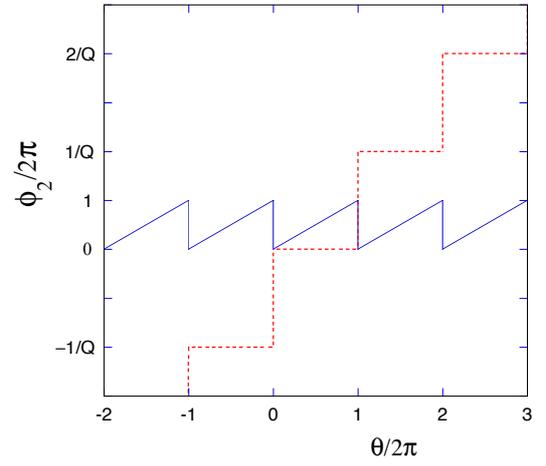}
\caption{ (Color online)
Phase $\phi_2$ of the second band as a function of the angle variable $\theta$
within the step-function approximation.
The dashed line indicates the phase difference $\phi_1-\phi_2$.
$\phi_2$ changes at $\theta=2\pi k$ for integer $k$.
$Q$ is the vorticity of a vortex.  The scale of vertical axis corresponds
to $Q=1/2$.
}
\label{phase2}
\end{center}
\end{figure}

\begin{figure}[htbp]
\begin{center}
\includegraphics[height=6.5cm]{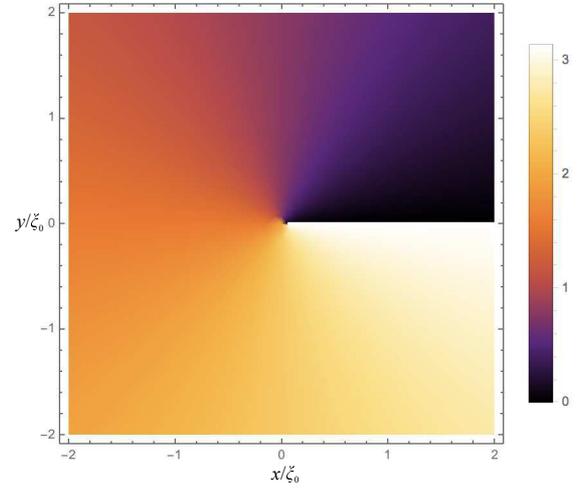}
\caption{ (Color online)
Phase $Q\phi_2=\phi_2/2$ for $Q=1/2$ in the second layer as a 
function of two-dimensional coordinates $x$ and $y$.
There is a kink on the positive part of $x$ axis. 
}
\label{phase3}
\end{center}
\end{figure}

\begin{figure}[htbp]
\begin{center}
\includegraphics[height=6.0cm]{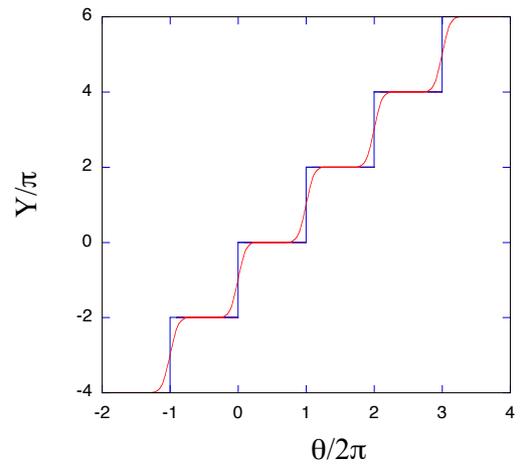}
\caption{ (Color online)
$Y(\theta)$ as a function of $\theta$ for $Q=1/2$.
$Y(\theta)$ is given by a step function in the step-function
approximation for the kink.
In general, $Y(\theta)$ is a smooth function of $\theta$ as
shown by the curve in the figure.
}
\label{phase4}
\end{center}
\end{figure}

\section{Wave Function of Variable Separation Form}

We will find a solution of the variable separation form:
\begin{equation}
f(r,\theta)= \exp\left( iY(\theta)\hat{C}\right)g(r),
\end{equation}
where $\hat{C}$ is a constant matrix, $Y(\theta)$ is a function of $\theta$
and $g(r)$ is a radial function of $r=|{\bf r}|$. 
The equation for $g(r)$ and $Y(\theta)$ reads
\begin{eqnarray}
&&\sigma_3\frac{\hbar^2}{2m}\Big[ -\frac{\partial^2 g}{\partial r^2}
-\frac{1}{r}\frac{\partial g}{\partial r}
-\left(iY''(\theta)\hat{C}-Y'(\theta)^2 K^2\right)\frac{1}{r^2}g-k_F^2g \nonumber\\
&& +\left(\mu-\frac{Q}{2}\sigma_3\right)^2\frac{1}{r^2}
\left(\frac{\partial\phi}{\partial\theta}\right)^2g
-i\left(\mu-\frac{Q}{2}\sigma_3\right)\frac{1}{r^2}
\frac{\partial^2\phi}{\partial\theta^2}g \nonumber\\
&& -2i\left(\mu-\frac{Q}{2}\sigma_3\right)iY'(\theta)\hat{C}
\frac{1}{r^2}\frac{\partial\phi}{\partial\theta}g
 \Big]
 +e^{-iY\hat{C}}\sigma_1e^{iY\hat{C}}\Delta_0(r)g \nonumber\\
&& = \epsilon_{\mu n} g,
\end{eqnarray}
where we assume that $\hat{C}$ commutes with $\sigma_3$.
We have a solution when $g(r)$ and $Y(\theta)$ satisfy the
following equations:
\begin{equation}
\sigma_3\frac{\hbar^2}{2m}\Big[ -\frac{\partial^2 g}{\partial r^2}
-\frac{1}{r}\frac{\partial g}{\partial r}
+\hat{M}\frac{1}{r^2}g-k_F^2g\Big]+e^{-iY\hat{C}}\sigma_1e^{iY\hat{C}}\Delta_0(r)g
= \epsilon_{\mu n}g,
\end{equation}
\begin{eqnarray}
&&-\left( iY''(\theta)\hat{C}-Y'(\theta)^2\hat{C}^2\right)
+\left(\mu-\frac{Q}{2}\sigma_3\right)^2
\left(\frac{\partial\phi}{\partial\theta}\right)^2 \nonumber\\
&&-i\left(\mu-\frac{Q}{2}\sigma_3\right)
\frac{\partial^2\phi}{\partial\theta^2}
 +2\left(\mu-\frac{Q}{2}\sigma_3\right)\hat{C}Y'(\theta)
\frac{\partial\phi}{\partial\theta} =\hat{M},
\nonumber\\
\end{eqnarray}
where $\hat{M}$ is a matrix depending on $\mu$ and possibly depends on $\theta$.
When there is no kink, we have $\phi(\theta)=\theta$ and $Y(\theta)=0$, and
then $\hat{M}= (\mu-Q\sigma_3/2)^2$. 
We will find a solution under the condition that $Y=0$ when $\phi(\theta)=\theta$.
$Y(\theta)$ may be expanded in terms of $\phi-\theta$ as:
$Y= a_1(\phi-\theta)+a_2(\phi-\theta)^2+\cdots$.
Since we adopt the step function approximation for $\phi$, $Y$ should be linear
in $\phi-\theta$ given as
\begin{equation}
Y(\theta)= \phi(\theta)-\theta,
\end{equation}
so that the delta-function singularity should be removed.
Then we obtain
\begin{align}
\hat{C}& = -\left( \mu-\frac{Q}{2}\sigma_3 \right),\\
\hat{M}&= \hat{C}^2 = \left( \mu-\frac{Q}{2}\sigma_3 \right)^2.
\end{align}
The equation for $g(r)$ is also followed with this matrix $\hat{M}$.
When $\phi(\theta)$ is given by the step-function approximation,
$Y(\theta)$ is given by a step function whose derivative has the form
\begin{equation}
Y'(\theta)= 2\pi\sum_{k\in {\bf Z}}\delta(\theta-2\pi k),
\end{equation}
where $k$ takes values for all integers.
The behavior of $Y(\theta)$ is shown in Fig. 4 where $Y(\theta)$
is shown as a function of $\theta$.

The angular dependence of $f(r,\theta)$ is given as
\begin{equation}
e^{iY(\theta)\hat{C}}
= e^{-iY(\theta)\mu}\left(
\begin{array}{cc}
e^{iQY(\theta)/2}  &  0 \\
0  &  e^{-iQY(\theta)/2}  \\
\end{array}
\right).
\end{equation}
When $Q=1/2$, this factor reads
\begin{equation}
e^{iY(\theta)\hat{C}}= e^{-iY(\theta)(\mu-1/4)}\left(
\begin{array}{cc}
1  &  0 \\
0  &  e^{-iY(\theta)/2}  \\
\end{array}
\right).
\end{equation}
Hence turning around the vortex core in the counterclockwise direction
changes the relative phase of $u$ and $v$ as
\begin{equation}
\left(
\begin{array}{cc}
\tilde{u}_{\mu n} \\
\tilde{v}_{\mu n} \\
\end{array}
\right) \rightarrow \left(
\begin{array}{cc}
-\tilde{u}_{\mu n} \\
\tilde{v}_{\mu n} \\
\end{array}
\right),
\end{equation}
since $Y(\theta)$ changes by $2\pi$.
When the quasiparticle goes around the vortex twice, $\tilde{u}_{\mu n}$
and $\tilde{v}_{\mu n}$ return to their original values.

Within the step-function approximation
the equations for
$^t g(r)\equiv (\tilde{u}_{\mu n}(r),\tilde{v}_{\mu n}(r))$ are
\begin{eqnarray}
&&\frac{d^2\tilde{u}_{\mu n}}{dr^2}+\frac{1}{r}\frac{d\tilde{u}_{\mu n}}{dr}
+k_F^2\tilde{u}_{\mu n}
-\frac{1}{r^2}\left( \mu-\frac{Q}{2}\right)^2\tilde{u}_{\mu n} 
 -\frac{2m}{\hbar^2}\Delta_0(r)\tilde{v}_{\mu n} \nonumber\\
&& ~~~~= -\frac{2m}{\hbar^2}\varepsilon_{\mu n} \tilde{u}_{\mu n}, \\
&&\frac{d^2\tilde{v}_{\mu n}}{dr^2}+\frac{1}{r}\frac{d\tilde{v}_{\mu n}}{dr}
+k_F^2\tilde{v}_{\mu n}
-\frac{1}{r^2}\left( \mu+\frac{Q}{2}\right)^2\tilde{v}_{\mu n} 
 +\frac{2m}{\hbar^2}\Delta_0(r)\tilde{u}_{\mu n} \nonumber\\
&& ~~~~= \frac{2m}{\hbar^2}\varepsilon_{\mu n} \tilde{v}_{\mu n},
\end{eqnarray}
where $\tilde{u}_{\mu n}(r)$ and $\tilde{v}_{\mu n}(r)$ are a radial function depending
only on $r$ (where we used the same symbols as $\tilde{u}_{\mu n}({\bf r})$ and
$\tilde{v}_{\mu n}({\bf r})$).

\section{Absence of Low-Lying Andreev Bound States in the Vortex Core}

We now examine the quasiparticle excitation spectra for an HFQV.
From the boundary condition for $\Psi({\bf r})=\Psi(r,\theta)$ given as
\begin{equation}
\Psi(r,\theta+2\pi/Q)= \Psi(r,\theta),
\end{equation}
the value of $\mu$ is quantized as
\begin{equation}
\mu= Q(\ell+1/2) ~~~ {\rm for}~\ell\in {\bf Z}.
\end{equation}
When we follow the conventional method\cite{car64,gen66,vol99}, the eigenvalue 
$\varepsilon$
is given as
\begin{equation}
\varepsilon_{\mu} = Q\mu\hbar\omega_0 = \left( \ell+1/2\right)Q^2\hbar\omega_0,
\end{equation}
where $\hbar\omega_0\simeq \Delta_{\infty}^2/\epsilon_F$.
The quasiparticle spectrum, however, is completely different from
this prediction for an FFQV.  
We show numerical results of excitation modes for an HFQV ($Q=1/2$)
in Fig. 5 and Fig. 6, where
length and energy are measured in units of $\xi_0$
and $\Delta_{\infty}$, respectively.
One notices a gap in excitation modes of an HFQV in the vicinity of
$\mu=0$.  This indicates that the Andreev bound state does not exist
near zero energy in the 
HFQV vortex core, in contrast to excitation modes of a vortex
with $Q=1$.

The spectra in Fig. 5 and Fig. 6 indicate a very surprising feature
for an HFQV.
The positive-energy quasi-particle state with negative angular momentum $L_z$ 
exists near $\mu=0$.
This indicates that
the negative-$L_z$ quasiparticle is moving in the opposite direction compared
with that with positive $L_z$.
The eigenfunctions of the Bogoliubov equation are shown in
Fig. 7 and Fig. 8, where we choose $\mu=1/4$ ($\ell=0$) (Fig. 7) and $\mu=-1/4$
($\ell=-1$) (Fig. 8).
We selected two states near $\mu=0$; the index $n=0$ indicates the highest negative
energy state and $n=1$ the lowest positive energy one, respectively. 
The wave function of the positive energy state with negative angular momentum 
shows the similar
behavior as that with the positive angular momentum state.

\begin{figure}[htbp]
\begin{center}
\includegraphics[height=6.5cm]{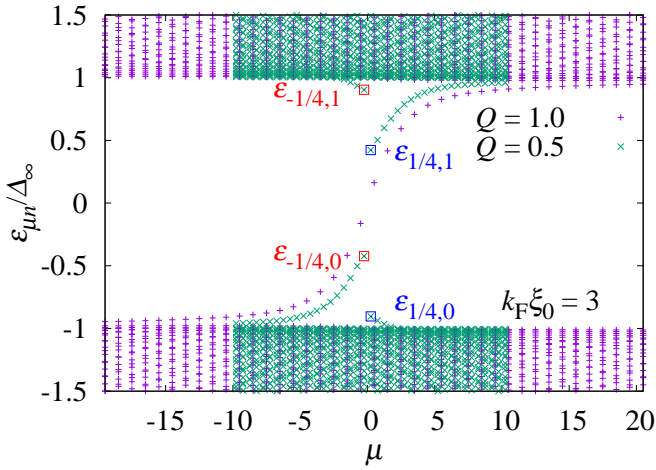}
\caption{ (Color online)
Excitation modes around an HFQV for $k_F\xi_0=3.0$.
The cutoff distance of a superconducting disk is $r_c=25\xi_0$.
}
\label{spec1}
\end{center}
\end{figure}

\begin{figure}[htbp]
\begin{center}
\includegraphics[height=6.5cm]{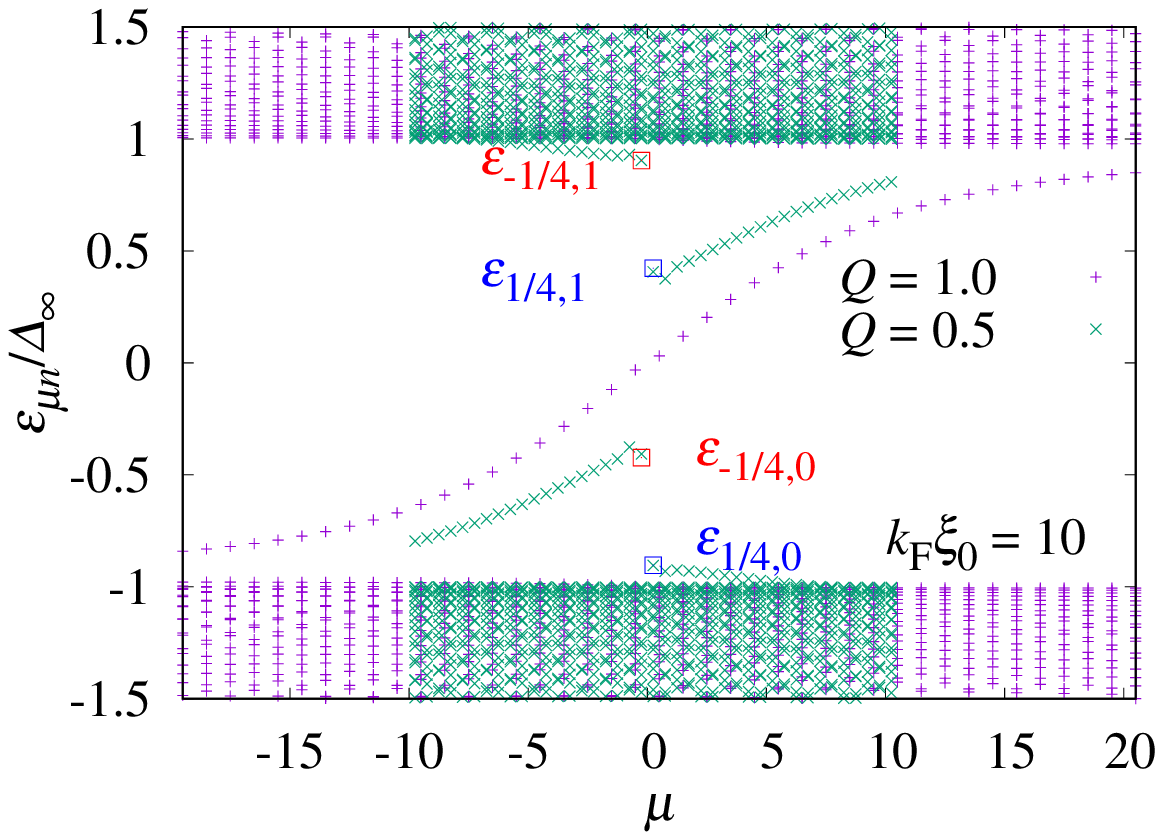}
\caption{ (Color online)
Excitation modes around an HFQV for $k_F\xi_0=10.0$.
The cutoff distance of a superconducting disk is $r_c=25\xi_0$.
}
\label{spec2}
\end{center}
\end{figure}

\begin{figure}[htbp]
\begin{center}
\includegraphics[height=6.5cm]{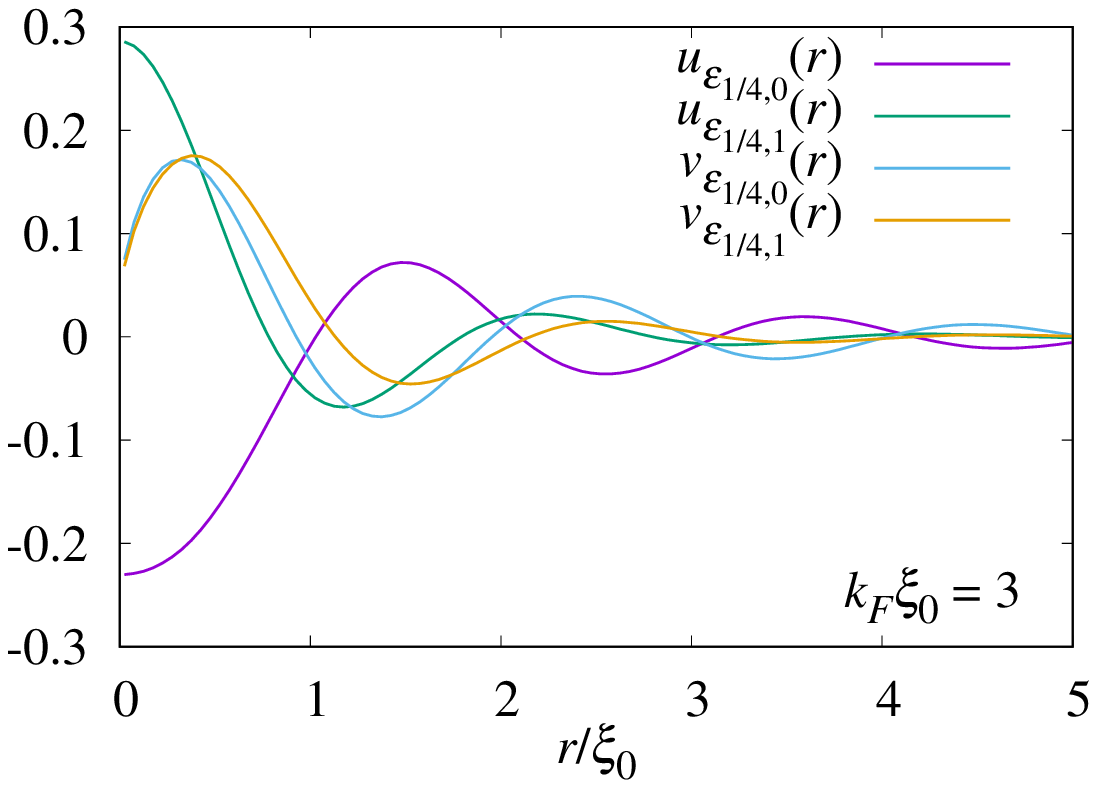}
\caption{ (Color online)
Eigenfunctions of the Bogoliubov equation for an HFQV for $\mu=1/4$
with $n=0$ and 1 where
 $k_F\xi_0=3.0$ and
the cutoff distance of an SC disk is $r_c=25\xi_0$.
The wave function for $\varepsilon_{1/4,0}$ corresponds to the
anomalous negative energy state with positive angular momentum.
}
\label{wf125}
\end{center}
\end{figure}

\begin{figure}[htbp]
\begin{center}
\includegraphics[height=6.5cm]{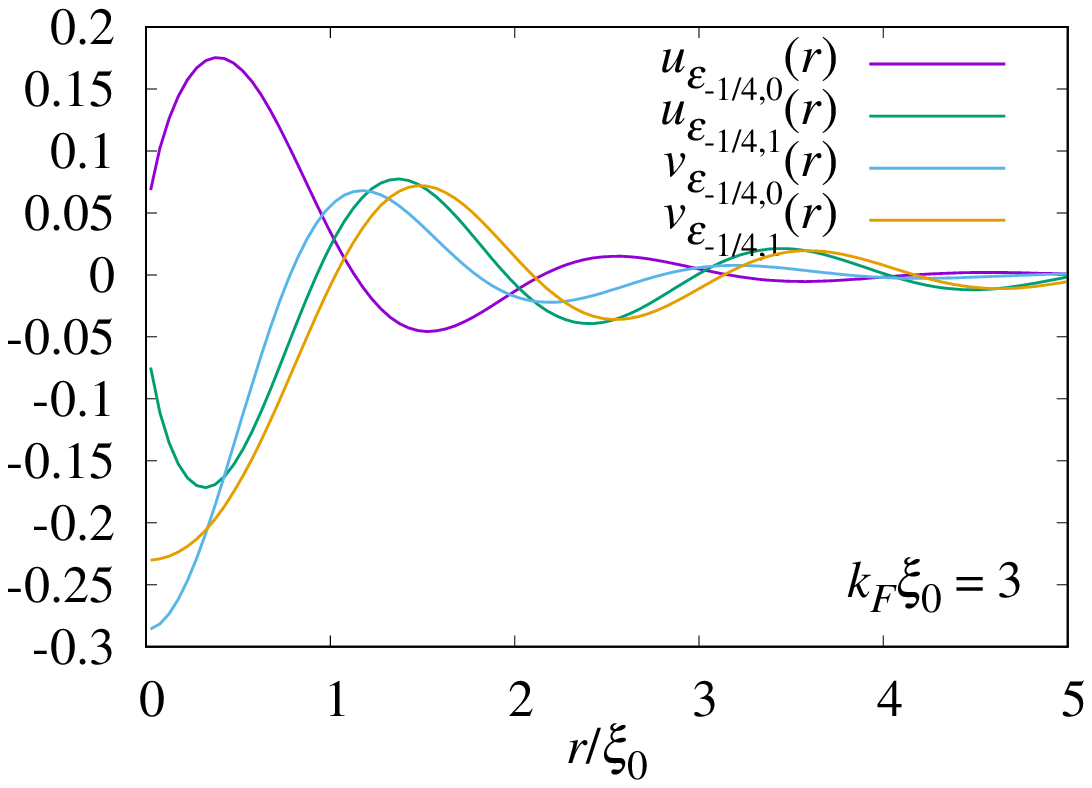}
\caption{ (Color online)
Eigenfunctions of the Bogoliubov equation for an HFQV for
$\mu=-1/4$ with $n=0$ and 1 where 
$k_F\xi_0=3.0$ and
the cutoff distance of an SC disk is $r_c=25\xi_0$.
The wave function for $\varepsilon_{-1/4,1}$ corresponds to the
anomalous positive energy state with negative angular momentum.
}
\label{wf625}
\end{center}
\end{figure}

{\em Phase change across the kink}
When $Y(\theta)$ is a general function of $\theta$,
the equation for the radial function $g(r)$ is written as
\begin{eqnarray}
&&\sigma_3\frac{\hbar^2}{2m}\Bigg[ -\frac{\partial^2 g}{\partial r^2}
-\frac{1}{r}\frac{\partial g}{\partial r}
+\hat{M}\frac{1}{r^2}g-k_F^2g\Bigg] 
+\left(
\begin{array}{cc}
0  &  e^{-iYQ} \\
e^{iYQ}  &  0  \\
\end{array}
\right)
\nonumber\\
&& ~~~ \times \Delta_0(r)g
 = \epsilon_{\mu n}g.
\end{eqnarray}
When $Y=0$, the potential term equals $\Delta_0(r)g$, while the
potential term becomes $-\Delta_0(r) g$ when $Y=2\pi$ with the
change of sign for $Q=1/2$.
As $Y$ changes from 0 to $2\pi$, the phase of $\tilde{u}_{\mu n}$
changes by $\pi$ and
and that of $\tilde{v}_{\mu n}$ remains the same.
This is shown in Figs. 9 and 10.
Thus the wave function is defined on a Riemann surface where two
planes are connected on the kink as the Riemann surface of 
complex function $w=z^{1/2}$ with $z=x+iy$.

\begin{figure}[htbp]
\begin{center}
\includegraphics[height=6.5cm]{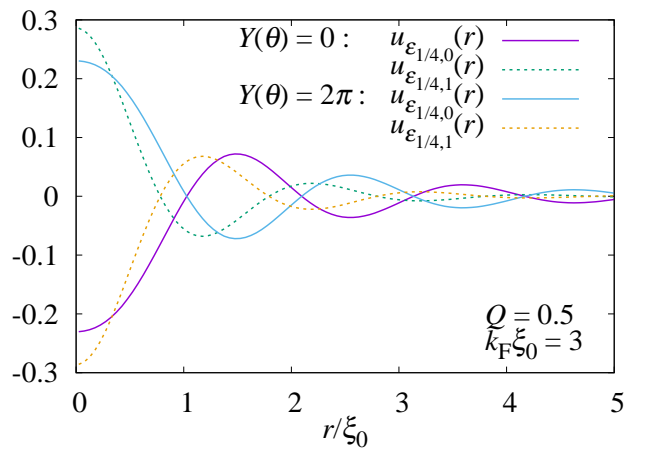}
\caption{ (Color online)
Eigenfunction $u$ of the Bogoliubov equation for an HFQV for $Y=0$ and $Y=2\pi$
with $\mu=1/4$, $n=0$ and 1 where
 $k_F\xi_0=3.0$ and
the cutoff distance of an SC disk is $r_c=25\xi_0$.
}
\label{wfu0-2pi}
\end{center}
\end{figure}

\begin{figure}[htbp]
\begin{center}
\includegraphics[height=6.5cm]{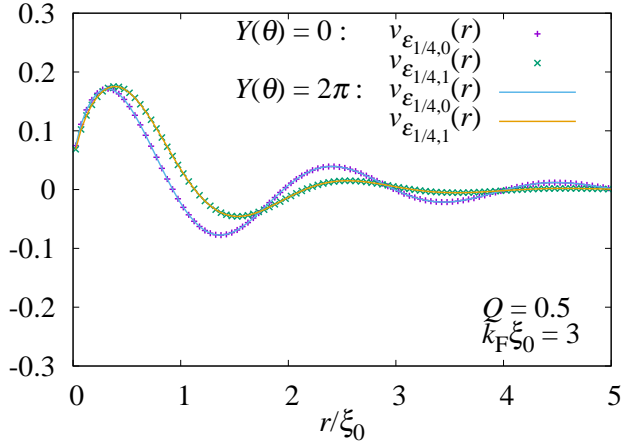}
\caption{ (Color online)
Eigenfunction $v$ of the Bogoliubov equation for an HFQV for $Y=0$ and $Y=2\pi$
with $\mu=1/4$, $n=0$ and 1 where
 $k_F\xi_0=3.0$ and
the cutoff distance of an SC disk is $r_c=25\xi_0$.
}
\label{wfv0-2pi}
\end{center}
\end{figure}

\section{Topological Index with Boundary and Fractional Skyrmion Number}

{\em Topological index with boundary}
The skyrmion number (or Chern number) is the topological number
that can be assigned to a vortex.
Usually this number is an integer for a vortex with integer vorticity,
which is related with property that the Andreev bound state exists
near zero energy.
We show that the topological numer is fractional for an FFQV. 
We write the Hamiltonian in the form,
\begin{align}
H&= \epsilon(x,y)\sigma_3+\Delta_0(r)\cos(Q\phi)\sigma_1
+\Delta_0(r)\sin(Q\phi)\sigma_2 \nonumber\\
&= E(x,y)\left( n_3\sigma_3+n_1\sigma_1+n_2\sigma_2 \right),
\end{align}
where $E(x,y)=\sqrt{\epsilon(x,y)^2+\Delta_0(r)^2}$.
${\bf n}\equiv (n_1,n_2,n_3)$ is a vector of unit length: $n_1^2+n_2^2+n_3^2=1$.
Immediately we can define the topological number associated with ${\bf n}$,
\begin{equation}
N \equiv \frac{1}{4\pi}\int {\bf n}\cdot\left( 
\frac{\partial {\bf n}}{\partial x}\times \frac{\partial {\bf n}}{\partial y}\right)
dxdy.
\end{equation}
Whether $N$ is an integer or not is dependent on the boundary condition
for $(\theta,Q\phi)$.
Let us consider an HFQV with $Q=1/2$.  The angle variable $\theta$ is
given by $\theta=\tan^{-1}(x/y)$, and when $\theta$ varies from 0 to $2\pi$, 
$Q\phi$ varies from 0 to $\pi$. 
$Q\phi=\phi/2$ is parametrized as
\begin{align}
\sin\left(Q\phi\right)&= \frac{1}{2i}\left( \left(\frac{z}{r}\right)^{Q}
-\left(\frac{\bar{z}}{r}\right)^{Q}\right),
\\
\cos\left(Q \phi\right)&= \frac{1}{2}\left( \left(\frac{z}{r}\right)^{Q}
+\left(\frac{\bar{z}}{r}\right)^{Q}\right),
\end{align}
for $z=x+iy$ and its complex conjugate $\bar{z}=x-iy$.
When the point $(x,y)$ moves on the entire plane for $0<\theta <2\pi$, 
$(\theta,Q\phi)$ covers
half of the sphere $S^2$. 

For an HFQV,
we have two contributions $N=N_{vortex}+N_{kink}$ from the regions 
$R= \{ \delta <\theta <2\pi-\delta\}$
and $\partial R=\{-\delta<\theta<\delta\}$ for small positive $\delta$,
respectively.  We can write $N_{vortex}=N_{bulk}$ and $N_{kink}=N_{boundary}$.
We define the angle $\varphi$ by
$\cos\varphi= \epsilon/\sqrt{\epsilon^2+\Delta_0^2}$ and
$\sin\varphi= \Delta_0/\sqrt{\epsilon^2+\Delta_0^2}$.  Then we have
\begin{align}
N_{vortex}&= \frac{Q}{4\pi}\int_{\theta\in R}\sin\varphi
\left( \frac{\partial \varphi}{\partial x}\frac{\partial\phi}{\partial y}
- \frac{\partial \varphi}{\partial y}\frac{\partial\phi}{\partial x}\right)
dx \wedge dy \nonumber\\
&= \frac{Q}{2}\int \frac{\Delta_0^2}{(\epsilon^2+\Delta_0^2)^{3/2}}d\epsilon
= Q =\frac{1}{2},
\end{align}
where we assume that $\epsilon$ takes the value $-\infty<\epsilon<\infty$
and $d\epsilon/dr\geq 0$.
$N_{vortex}$ is regarded as the skyrmion number for the vortex.
When $\theta\in \partial R$, we use $d\phi/d\theta=2\pi \delta(\theta)$
so that we obtain
\begin{equation}
N_{kink}= \frac{Q}{4\pi}\int_{\theta\in\partial R}\sin\varphi
\frac{\partial\varphi}{\partial r} 2\pi\delta(\theta)dr\wedge d\theta
= Q = \frac{1}{2}.
\end{equation}
The total index $N$ is given by 
\begin{equation}
N=N_{vortex}+N_{kink}=1.
\end{equation}
This indicates that the topological number is divided into two
contributions from the vortex and the kink (boundary).
This is the reason why the vortex has the fractional topological number.
This may be regarded as a kind of the index theorem for manifolds with 
boundary\cite{ati75,mel93,fuk17}.

{\em Skyrmion on a Riemann surface}
This is generalized for general rational $Q$, where we have
\begin{equation}
N_{vortex}=Q.
\end{equation}
Thus the topological number $N_{vortex}$ is a fractional number for an FFQV.
An FFQV can be regarded as a skyrmion on a Riemann surface.
For an HFQV, $\phi$ is defined  on the Riemann surface defined by the
function $w=z^{1/2}$.  
The Riemann surface $w=z^{1/2}$ is identical to $S^2$ by compactification, which
induces a map $S^2\rightarrow S^2$.
The topological number $N_{vortex}$ is the winding number of this map.
Thus $N_{vortex}$ calculated on the
Riemann surface is $N_{vortex}=1$.  $N_{vortex}$ for an HFQV corresponds to
the integration on one sheet and this results in $N_{vortex}=1/2$.

{\em Non-abelian statistics of fractional-quantum vortices}
We argue that fractional-quantum vortices follow non-abelian statistics.
Suppose that an FFQV goes around the other FFQV in the counterclockwise direction.
The phase of an electron and a hole in the FFQV changes $Q\pi$ and
$-Q\pi$, respectively.  For an HFQV, the wave function takes phase factors
$i$ and $-i$ for electrons and holes, respectively.
Then the Bogoliubov operator is transformed to the other operator
in the process of exchange of HFQVs and the Bogoliubov amplitudes are
transformed as
\begin{eqnarray}
\left(
\begin{array}{c}
u \\
v \\
\end{array}
\right) \rightarrow \left(
\begin{array}{c}
u \\
-v \\
\end{array}
\right).
\end{eqnarray}
This indicates that two HFQVs follow non-abelian statistics.
This differs from the non-abelian statistics in p-wave 
superconductors\cite{iva01}.

\section{Discussion}

We have investigated quasi-particle excitation modes in a half-flux
quantum vortex by solving the Bogoliubov equation numerically.
We have found that there is no low-lying Andreev bound state near zero
energy in the vortex core, that is, there is the energy region
where no Andreev bound states exist.
The skyrmion number $N_{vortex}$ becomes fractional for an FFQV.
An HFQV is nothing but a half-skyrmion.
The index $N_{kink}$ is just the boundary contribution, and
$N_{vortex}=N_{bulk}$ becomes fractional due to this:
$N_{vortex}= N-N_{kink}$.

We discuss scanning tunneling microscope (STM) observations
of HFQV here.  
It is important that the existence of HFQV will be confirmed by
measurements by STM measurements because the quasiparticle spectra are different
between HFQVs and conventional vortices.
The phase of $\tilde{u}$ takes two values 0 or $\pi$ 
depending on which side of the Riemann surface the quasiparticle lies.
The density of states depends on $|\tilde{u}|^2$
and $|\tilde{v}|^2$ and is independent of the phase of $\tilde{u}$.
Thus scanning tunneling spectroscopy (STS) results will not depend on 
the plane of the
Riemann surface of the phase of $\tilde{u}$.

{\em Acknowledgment}
This work was supported in part by Grant-in-Aid from the
Ministry of Education, Culture, Sports and Science
(MEXT) of Japan (No. 17K05559).


\begin{thebibliography}{9}

\bibitem{mos59} V. A. Moskalenko: Fiz. Metal and Metallored 8, 2518
(1959).
\bibitem{suh59} H. Suhl, B. T. Mattis and L. W. Walker: Phys. Rev. Lett.
3, 552 (1959).
\bibitem{per62} J. Peretti, Phys. Lett. 2, 275 (1962).
\bibitem{kon63}J. Kondo: Prog. Theor. Phys. 29, 1 (1963).

\bibitem{sta10}
V. Stanev and Z. Tesanovic, Phys. Rev. B81, 134522 (2010).
\bibitem{tan10a}
Y. Tanaka and T. Yanagisawa,
J. Phys. Soc. Jpn. 79, 114706 (2010).
\bibitem{tan10b}
Y. Tanaka and T. Yanagisawa, Solid State Commun. 150, 1980 (2010).
\bibitem{dia11}
R. G. Dias and A. M. Marques, Supercond. Sci. Technol. 24, 085009 (2011).
\bibitem{yan12}T. Yanagisawa, Y. Tanaka, I. Hase and K. Yamaji, J. Phys. Soc. Jpn.
81, 024712 (2012).
\bibitem{hu12}X. Hu and Z. Wang, Phys. Rev. B85, 064516 (2012).
\bibitem{sta12}V. Stanev, Phys. Rev. B85, 174520 (2012).
\bibitem{pla12}C. Platt, R. Thomale, C. Homerkamp and S. C. Zhang,
Phys. Rev. B85, 180502 (2012).
\bibitem{mai13}S. Maiti and A. V. Chubukov, Phys. Rev. B87, 144511 (2013).
\bibitem{wil13}B. J. Wilson and M. P. Das,
J. Phys. Condens. Matter 25, 425702 (2013).
\bibitem{gan14} R. Ganesh, G. baskaran, J. van den Brink and D. V. Efremov,
Phys. Rev. Lett. 113, 177001 (2014).
\bibitem{yer15} Y. S. Yerin, A. N. Omelyanchouk and E. Il'ichev,
Super. Sci. Technol. 28, 095006 (2015).

\bibitem{yan13}T. Yanagisawa and I. Hase,
J. Phys. Soc. Jpn. 82, 124704 (2013).
\bibitem{lin12}S. Z. Lin and X. Hu, New J. Phys. 14, 063021 (2012).
\bibitem{kob13}K. Kobayashi, Y. Ota, M. Machida and H. Aoki,
Phys. Rev. B88, 224516 (2013).
\bibitem{koy14} T. Koyama, J. Phys. Soc. Jpn. 83, 074715 (2014).
\bibitem{yan14}T. Yanagisawa and Y. Tanaka, New J. Phys. 16, 123014 (2014).
\bibitem{tan15} Y. Tanaka, I. Hase, T. Yanagisawa, G. Kato, T. Nishio and
S. Arisawa, Physica C516, 10 (2015).
\bibitem{sha02} S. G. Sharapov, V. P. Gusynin  and H. Beck,
Eur. Phys. J. B39, 062001 (2002).
\bibitem{yan17c} T. Yanagisawa, J. Phys. Soc. Jpn. 86, 104711 (2017).

\bibitem{izy90} Yu. A. Izyumov and V. M. Laptev, Phase Transitions
20, 95 (1990).
\bibitem{vol09} G. E. Volovik, {\em The Universe in a Helium Droplet}
(Oxford University Press, Oxford, 2009).
\bibitem{tan02} Y. Tanaka: Phys. Rev. Lett. 88, 017002 (2001).
\bibitem{bab02} E. Babaev, Phys. Rev. Lett. 89, 067001 (2002).
\bibitem{col05} A. D. Col, V. B. Geshkenbein, and G. Blatter,
Phys. Rev. Lett. 94, 097001 (2005).
\bibitem{blu06} H. Bluhm, N. C. Koshnick, M. E. Huber and
K. A. Moler, Phys. Rev. Lett. 97, 237002 (2006).
\bibitem{gor07} J. Goryo, S. Soma, and H. Matsukawa, Europhys. Lett.
80, 17002 (2007).
\bibitem{chi10} L. F. Chibotaru and V. H. Dao, Phys, Rev. B81, 020502
(2010).
\bibitem{kup11} S. V. Kuplevakhsky, A. N. Omelyanchouk and Y. S. Yerin,
Low Temp. Phys. 37, 667 (2011).
\bibitem{gar11} J. Garaud, J. Carlstrom and E. Babaev,
Phys. Rev. Lett. 107, 197001 (2011).
\bibitem{pin12} J. C. Pina, C. C. de Souza Silva, and M. V.
Miloevic, Phys. Rev. B86, 024512 (2012).

\bibitem{gar13} J. Garaud, J. Carlstrom, E. Babaev and M. Speight,
Phys. Rev. B87, 014507 (2013).
\bibitem{smi05} J. Smiseth, E. Smorgrav, E. Babaev and A. Sudbo,
Phys. Rev. B71, 214509 (2005).

\bibitem{tan18} Y. Tanaka, H. Yamamori, T. Yanagisawa, T. Nishio,
S. Arisawa,
Physica C548, 44 (2018).
\bibitem{hay19} M. Hayashi, J. Phys. Soc. Jpn. 88, 035002 (2019).
\bibitem{nag18} Y. Nagai and Y. Kato, J. Phys. Soc. Jpn. 88, 054707 (2019).

\bibitem{cho09} H. Y. Choi, J. H. Yun, Y. Bang and H. C. Lee, 
Phys. Rev. B80, 052505 (2009).
\bibitem{shi09} P. M. Shirage, K. Kihou, K. Miyazawa, C.-H. Lee,
H. Kito, H. Eisaki, T. Yanagisawa, Y. Tanaka and A. Iyo, Phys. Rev. Lett. 
103, 257003 (2009).
\bibitem{yan09} T. Yanagisawa, K. Odagiri, I. Hase, K. Yamaji,
P. M. Shirage, Y. Tanaka, A. Iyo and H. Eisaki, J. Phys. Soc. Jpn. 78, 094718 (2009).
\bibitem{vir99} S. M. M. Virtanen and M. M. Salomaa, Phys. Rev.
B60, 14581 (1999).

\bibitem{yan18} T. Yanagisawa, I. Hase, and Y. Tanaka, Phys. Lett. 
A382, 3483 (2018).

\bibitem{car64} C. Caroli, P. G. de Gennes and J. Matricon,
Phys. Lett. 9, 307 (1964).
\bibitem{gen66} P. G. de Gennes, {\em Superconductivity of
Metals and Alloys} (W. A. Benjamin, New York, 1966).
\bibitem{vol99} G. E. Volovik, JETP Lett. 70, 609 (1999).
\bibitem{kul70} I. O. Kulik, Sov. Phys. JETP 30, 844 (1970).
\bibitem{sto96} M. Stone, Phys. Rev. B54, 13222 (1996).

\bibitem{ati75} M. F. Atiyah, V. K. Patodi, and I. M. Singer,
Math. Proc. Cambridge Philos. Soc. 77, 43 (1975).
\bibitem{mel93} R. B. Melrose, {\em The Atiyah-Patodi-SInger} Index
Theorem (Taylor \& Francis, Milton Park, UK, 1993).
\bibitem{fuk17} H. Fukaya, T. Onogi, and S. Yamaguchi, Phys. Rev.
D 96, 125004 (2017).

\bibitem{iva01} D. A. Ivanov, Phys. Rev. Lett. 86, 268 (2001).

\end{thebibliography}
\end{document}